\documentclass[runningheads]{svmult}

\usepackage{amssymb}

\usepackage{makeidx}   
\usepackage{graphicx}  
\usepackage{subeqnar}  
\usepackage{multicol}  
\usepackage{physprbb}  
\makeindex             
\newcommand{\greeksym}[1]{{\usefont{U}{psy}{m}{n}#1}}
\newcommand{\umu}{\mbox{\greeksym{m}}}

\newcommand{\bdelta}{\mbox{\boldmath$\delta$}}
\newcommand{\bnabla}{\mbox{\boldmath$\nabla$}}
\newcommand{\bd}{{\mbox{\boldmath$d$}}}
\newcommand{\bk}{{\mbox{\boldmath$k$}}}
\newcommand{\br}{{\mbox{\boldmath$r$}}}
\newcommand{\bs}{{\mbox{\boldmath$s$}}}
\newcommand{\bu}{{\mbox{\boldmath$u$}}}
\newcommand{\bv}{{\mbox{\boldmath$v$}}}
\newcommand{\bA}{{\mbox{\boldmath$A$}}}
\newcommand{\bB}{{\mbox{\boldmath$B$}}}
\newcommand{\bD}{{\mbox{\boldmath$D$}}}
\newcommand{\bE}{{\mbox{\boldmath$E$}}}
\newcommand{\bI}{{\mbox{\boldmath$I$}}}
\newcommand{\bJ}{{\mbox{\boldmath$J$}}}
\newcommand{\bP}{{\mbox{\boldmath$P$}}}
\newcommand{\bS}{{\mbox{\boldmath$S$}}}
\newcommand{\bV}{{\mbox{\boldmath$V$}}}

\begin{document}
\title*{Numerical Simulations of Magnetic Fields 
        \protect\newline in Astrophysical Turbulence
        \protect\newline{\small\it Invited Review
                                   to be published in the Proceedings of the Workshop on
                                   Simulations of MHD Turbulence in Astrophysics, 
                                   Paris, France, July 2001, 
                                   eds. E.~Falgarone \& T.~Passot, Springer-Verlag.}}
\toctitle{Numerical Simulations of Magnetic Fields
\protect\newline in Astrophysical Turbulence}
%
%
\titlerunning{Simulations of Magnetic Fields}
%
\author{Ellen G. Zweibel\inst{1}
\and Fabian Heitsch\inst{1}
\and Yuhong Fan\inst{2}}
\authorrunning{Ellen G. Zweibel et al.}
%
%
\institute{JILA, University of Colorado, Boulder CO 80309, USA
\and High Altitude Observatory, NCAR, Boulder, CO 80307, USA}

\maketitle              

\begin{abstract}
The generation and evolution of astrophysical magnetic fields occurs
largely through the action of turbulence. In many situations, the magnetic 
field is strong enough to
influence many important properties of turbulence itself. Numerical simulation
of magnetized turbulence is especially challenging in the astrophysical
regime because of the high magnetic Reynolds numbers involved, but some aspects
of this difficulty can be avoided in weakly ionized systems.
\end{abstract}

\section{Introduction}

The interaction of magnetic fields with turbulence is a basic feature of many
astrophysical systems. 
Important, basic MHD turbulence problems common to many fields of
astrophysics include the
nature of MHD turbulence itself, the dynamo problem, the effects of magnetic
fields on turbulent transport and turbulent mixing, the effects of small
scale turbulence on large scale dynamics, and the formation and evolution
of current sheets.

Although progress on all of these problems has been made analytically,
numerical simulations are an increasingly powerful means of approaching them.
Accurate simulation of astrophysical magnetic fields under turbulent
conditions presents extreme challenges of its own. The main reason is
the smallness of the magnetic diffusivity, which leads naturally to the
formation of thin current layers. Most of the Ohmic dissipation, and most of
the change in magnetic topology, takes place in these layers. It is important
to understand these diffusive effects in order to answer questions such as:
How do dynamos amplify magnetic fields on large scales but not small scales?
What determines the ratio of magnetic flux to mass in self gravitating
regions? How is magnetic energy dissipated in a turbulent plasma?

The purpose of this review is to isolate some important problems related to
magnetized turbulence in astrophysics, with emphasis on their computational
aspects. The subject is vast, and we make no claim to be comprehensive. We
cite literature up to early 2002.

In \S 2, we write down the magnetic induction equation, derive some of its
basic properties, and discuss the use of conservation laws in testing MHD
codes. As an illustration, we carry out such tests on the ZEUS code. In \S 3,
we discuss the parameter space for astrophysical MHD. In \S 4, we discuss
some results on the dynamo problem, the turbulence problem, the formation of
singularities, and the role of turbulence in large scale dynamics. In \S 5, we
discuss lightly ionized media, for which ambipolar drift yields an effective
diffusivity which is much higher than the Ohmic value. Although the two forms
of diffusion are not equivalent, 
ambipolar drift is a partial solution to the problem of small
diffusivity. Section 6 is a summary and discussion of future prospects.

\section{The Magnetic Induction Equation: Theory and Tests}

\subsection{Induction Equation and Consequences}

According to
Faraday's law, also known as the magnetic
induction equation
\begin{equation}\label{faraday}
\frac{\partial\bB}{\partial t}=-c\bnabla\times\bE.
\end{equation}
Throughout this paper, we will assume $\bE$ is given by
\begin{equation}\label{E}
c\bE=-\bv\times\bB +\sigma^{-1}c\bJ=-\bv\times\bB +\lambda_{\Omega}\bnabla
\times\bB,
\end{equation}
where $\sigma$ is the electrical conductivity, $\lambda_{\Omega}\equiv c^2/4\pi
\sigma$ is the magnetic diffusivity, and in the last step we have used Ampere's
law, neglecting the displacement current. 

The first and second terms on the
RHS of eqn. (\ref{E}) represent inductive and resistive effects,
respectively.
In order to compare these terms, we
introduce a characteristic speed $V_0$ and a characteristic lengthscale $L_0$,
and write $\bv$ and $\br$ in terms of dimensionless velocities and coordinates;
$\bv\equiv V_0\bu$; $\br\equiv L_0\bs$. Equation (\ref{E}) can then be written
as
\begin{equation}\label{Eds}
c\bE=-V_0\left(\bu\times\bB - R_m^{-1}\bnabla_{\bs}\times\bB
\right),
\end{equation}
where the
dimensionless parameter $R_m$ is defined by
\begin{equation}\label{Rm}
R_m\equiv \frac{L_0V_0}{\lambda_{\Omega}}.
\end{equation} 
In some
problems it is convenient to set $V_0$ to 
a typical Alfv\'{e}n speed $v_A\equiv B/(4\pi
\rho)^{1/2}$, in which case $R_m$ is known as the Lundquist number and
denoted by $S$.
 
The case $\lambda_{\Omega}\propto S^{-1}\equiv 0$ is called \textit{ideal} 
MHD. It is clear
from eqns. (\ref{faraday}) and (\ref{E}) that the limit $S
\rightarrow\infty $ is a singular limit, in the sense that at this limit the order of
the magnetic induction equation drops from second to first.
 We therefore expect that as  $S\rightarrow\infty $, 
thin boundary layers, or current sheets, will form. This should be anticipated
when choosing numerical schemes (\S\ref{subsec:numimplications}).

The high $S$ limit
describes most astrophysical problems.
The ideal form of the magnetic induction equation can be solved exactly in
terms of fluid trajectories.
Suppose that the fluid at position $\br$ at time $t$ was at position
$\br_0$ at time $0$. Define the deformation matrix $\bD$ by
\begin{equation}\label{D}
D_{ij}=\frac{\partial r_i}{\partial r_{0j}}.
\end{equation}
It can then be shown that the magnetic field $\bB(\br,t)$ is related to
the initial field $\bB_0(\br_0(\br,t),0)$ by
\begin{equation}\label{Cauchy}
\bB(\br,t)=\frac{\bD\cdot\bB_0(\br_0(\br,t),0)}{\vert\bD\vert},
\end{equation}
where $\vert\bD\vert$ is the determinant of $\bD$.
Equation (\ref{Cauchy}) is known as the Cauchy solution of the magnetic
induction equation. It is the basis for the numerical technique used by
 Kinney et al.~\cite{KCC2000} to simulate 2D MHD turbulence.

Conservation laws can be used to test MHD codes.
The Cauchy solution embodies magnetic flux conservation. This 
basic property is in practice difficult to test, 
because in turbulent flow, fluid elements follow complex paths. 
The Cauchy
solution only makes sense as long as fluid trajectories do not
intersect, but this cannot be guaranteed in a finite difference code, which
is always somewhat diffusive. Therefore, we turn to a globally conserved
quantity: magnetic helicity.

The helicity $H$
of a fixed volume $V$ of fluid with magnetic vector potential $\bA$ and
 magnetic field $\bB=\bnabla\times\bA$ is
\begin{equation}\label{helicity}
H\equiv\int_Vd^3r\bA\cdot\bB.
\end{equation}
Uncurling eqn. (\ref{faraday}) leads to an evolution equation for $\bA$
\begin{equation}\label{uncurl}
\frac{\partial\bA}{\partial t}=-c\bE+\bnabla\phi,
\end{equation}
where $\phi$ is a free gauge function. According to eqns. (\ref{faraday}) and
(\ref{uncurl}), the rate of change of $H$ is
\begin{equation}\label{hdot}
\frac{dH}{dt}=-2c\int_Vd^3r\bE\cdot\bB + \int_Sd\bS\cdot\left(\bB\phi - c
\bE\times\bA\right),
\end{equation}
where we have integrated once by parts and used Gauss's theorem. We now
take periodic boundary conditions on $V$, so that
the surface integral vanishes, and use eqn. (\ref{E}).
Equation (\ref{hdot}) then becomes
\begin{equation}\label{hdot2}
\frac{dH}{dt}=-2\int_Vd^3r\lambda_{\Omega}\bB\cdot\bnabla\times\bB.
\end{equation}
Equation (\ref{hdot2}) shows that helicity is conserved in an ideal medium.
The rate at which helicity varies is a global
measure of the magnetic diffusivity $\lambda_{\Omega}$. Diffusion can cause
helicity growth as well as decay.

We obtain a slightly different conservation law if we assume that $V$ is
comoving instead of fixed. In this case, we find that in an ideal medium,
$H$ is conserved within $V$ as long as $\bB\cdot d\bS\equiv 0$. However,
in view of the difficulty of following comoving volumes in a turbulent fluid,
it is more useful to treat $V$ as fixed, and to take it as the computational
domain.

Finally, we derive an equation for
magnetic energy $W$
\begin{equation}\label{W}
W\equiv\int_Vd^3r\frac{B^2}{8\pi}.
\end{equation}
Taking the scalar product of eqn. (\ref{faraday}) with $\bB$, integrating over 
space, and assuming periodic boundary conditions yields
\begin{equation}\label{Wdot}
\frac{dW}{dt}=-\int_Vd^3r\bv\cdot\frac{\bJ\times\bB}{c} - 
\int_Vd^3r\lambda_{\Omega}\frac{\vert\bnabla\times\bB\vert^2}{4\pi}.
\end{equation}
The first term on the RHS of eqn. (\ref{Wdot}) represents the work done by the
flow on the field, and appears with opposite sign in the evolution equation
for kinetic energy.
The second term represents energy loss by Ohmic decay. 

\subsection{Helicity Conservation in the ZEUS Code}

The ZEUS-3D code \cite{SNA1992,SNB1992} solves the equations of ideal, 
compressible MHD using a finite difference scheme and a von Neumann artificial viscosity to
capture shocks. The MHD induction equation is followed using the method of consistent transport
along characteristics \cite{HAS1995}. 
ZEUS-3D is publically available, and has been of great service in the astrophysical community.

If there were no numerical
dissipation in the ZEUS code, magnetic helicity would be strictly conserved
(see eqn. (\ref{hdot2})).
Here, we investigate helicity conservation in the ZEUS code in two
applications.

The first problem is the evolution of a twisted magnetic flux
tube which is unstable to the kink mode. Initially, the helicity of the
tube is entirely in the twist of the field about the axis. Theory
predicts that the instability
drives the system to a new equilibrium state in which some of the helicity
is carried by a writhing deformation of the tube axis, and the total
magnetic energy is reduced while the total helicity is fixed.

\begin{figure}[h]
  \begin{center}
  \includegraphics[width=.9\textwidth]{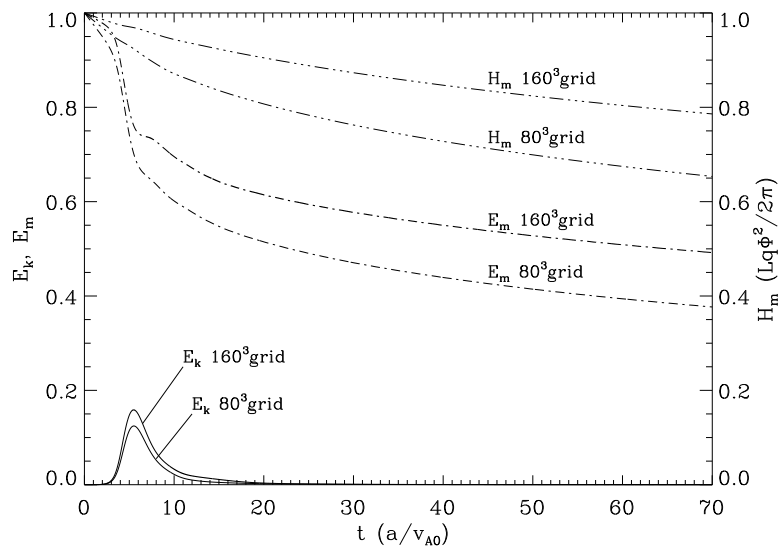}
  \end{center}
  \caption[]{Kinetic and magnetic energy, and helicity against time for the 
             twisted magnetic flux tube. $t$ is given in units of the Alfv\'{e}n crossing time
             for the flux tube with diameter $a$. Helicity is given in units of
 $Lq \Phi^2 /2 \pi$, where
$Lq / 2 \pi$ is the number of rotations by $2\pi$ of the twisted field lines 
about the tube axis over the domain length L, and $\Phi$ is the total
axial flux
of the tube.}

  \label{fig1}
\end{figure}

The simulations confirm the broad outline of this picture. The
growth rate of the kink during the linear phase agrees with an analytical 
calculation, and significant motion occurs only during
the kinking phase, while magnetic energy is being released. However, both
magnetic energy and magnetic helicity decline steadily once the system
reaches equilibrium, as shown in Figure (\ref{fig1}) for computations with
80$^3$ and 160$^3$ gridpoints.
During the dynamical kink phase, the energy declines much
faster than the helicity, confirming the importance of both dynamical and
resistive processes in the evolution of magnetic energy.

\begin{figure}[h]
  \begin{center}
  \includegraphics[width=.9\textwidth]{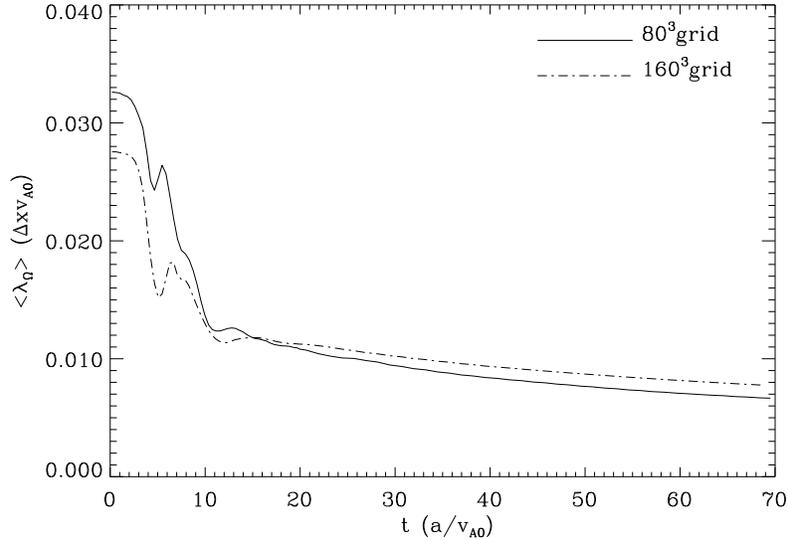}
  \end{center}
  \caption[]{Mean magnetic diffusivity against time for the twisted magnetic flux tube.}
  \label{fig2}
\end{figure}

We have used eqn. (\ref{hdot2}) to estimate the mean pseudo- magnetic 
diffusivity $\langle\lambda_{\Omega}\rangle$, i.e. 
the mean numerical magnetic diffusivity,
by writing the integral on the right hand side as
\begin{equation}\label{meaneta}
\int_Vd^3r\lambda_{\Omega}\bB\cdot\bnabla\times\bB = \langle\lambda_{\Omega}\rangle\int_Vd^3r\bB\cdot\bnabla\times\bB,
\end{equation}
which can be regarded as the definition of $\langle\lambda_{\Omega}\rangle$.
The result is shown in Figure (\ref{fig2}), where $\langle\lambda_{\Omega}\rangle$ is plotted for the two simulations in units of $\Delta x v_{A0}$, where
$\Delta x$ is the grid scale. The
near coincidence of the two curves shows that the numerical diffusion is
linear in $\Delta x$.
During the dynamical phase, $\langle\lambda_{\Omega}\rangle$ is enhanced
over its ``quiet" value $\epsilon_0\Delta x v_{A0}$
by about a factor of 3, perhaps suggesting that numerical
resistivity, like numerical viscosity, is proportional to velocity.

The results shown in Figure (\ref{fig2}) allow us to estimate the Lundquist
number $S$. According to eqn. (\ref{Rm}), $S$ can be written in terms of
$\epsilon_0$, the number of gridpoints $N$, and the magnetic lengthscale
$L_B$ in units of the box size $L$ as

\begin{equation}\label{Szeus}
S=\frac{L_B}{L}\frac{N^{1/3}}{\epsilon_0}.
\end{equation}
In these computations, the tube has an initially Gaussian magnetic profile with
$L_B/L = 0.1$. Taking $\epsilon_0\sim 0.01$, which is representative of the
``quiet" value shown in Figure (\ref{fig2}) and $N^{1/3}\sim 100$, we see
that $S\lesssim 10^3$.

\begin{figure}[h]
  \begin{center}
  \includegraphics[width=.7\textwidth]{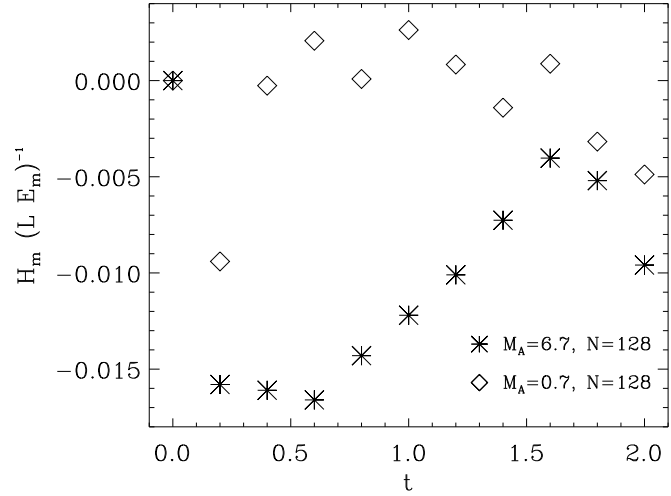}
  \end{center}
  \caption[]{Total magnetic helicity against time for two models of turbulent
             molecular clouds at $M_A = 0.7$ and $M_A = 6.7$.}
  \label{fig3}
\end{figure}

We have also computed the evolution of $H$ in models of molecular
clouds \cite{HMK2001,HZM2001}, with initially uniform magnetic fields, stirred by
supersonic turbulence. In these models $H$ is
initially zero, and should remain so, although the helicity density
$\bA\cdot \bB$ can vary arbitrarily
between positive and negative values from
point to point. Figure (\ref{fig3}) shows $H$ relative to $E_mL$ for
two different runs at various times, measured in units of the sound crossing time.
In these units, the errors in $H$ are at the 1\%
level, and, reassuringly, there is no evidence that $H$ drifts steadily
away from zero. The variance of $H$ around zero is smaller for the run at
Alfven Mach number 
$M_A = 0.7$ than for the run at $M_A = 6.7$, presumably because the field becomes
less tangled if it is relatively strong.

\begin{figure}[h]
  \begin{center}
  \includegraphics[width=.5\textwidth]{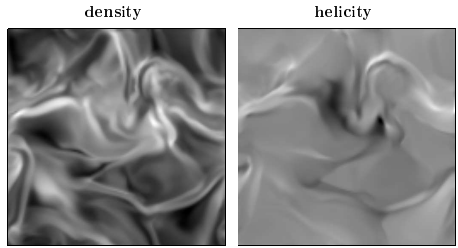}
  \end{center}
  \caption[]{Density (left panel) and helicity (right panel) for the weak field model
             with $M_A = 6.7$.}
  \label{fig4}
\end{figure}

Figure (\ref{fig4}) shows mass density and helicity density, for the
weak field model, in a plane
perpendicular to the mean magnetic field. The figure
shows that helicity density
varies strongly with
position, and that it is well correlated with mass density (the correlation
is not as good in the strong field case).

The density
compressions in these models are associated with shock fronts. 
Helicity density can also increase
across a shock. For example, consider a locally
sheared, force free magnetic field of the form
\begin{equation}\label{helical}
\bB=B_0\left(\hat x\cos{k_0z}+\hat y\sin{k_0z}\right).
\end{equation}
The vector potential is
\begin{equation}\label{vec}
\bA=-k_0^{-1}\bB,
\end{equation}
so the helicity density is $-k_0^{-1}B_0^2$. Suppose the fluid is compressed
in the $\hat z$ direction, so that the initial and final coordinates $z_0$
and $z$ are related by $z=\alpha z_0$, with $\alpha < 1$. It can be shown
from eqn. (\ref{Cauchy}), or from intuitive
arguments, that $\bB$ still has the form
(\ref{helical}), but with $B_0\rightarrow\alpha^{-1}B_0$ and $k_0
\rightarrow\alpha^{-1}k_0$. The helicity density then increases by a factor
$\alpha^{-1}$, which is the same factor by which the mass density increases.

\section{Parameter Space}

The dimensionless parameters of astrophysical turbulence define the scope
of the problem and should influence the choice of numerical technique. Here
we define and give quantitative expressions for some important parameters.

\subsection{Macroscopic Parameters\label{subsubsec:macroparm}}

The ratio of gas pressure to magnetic pressure is usually denoted by $\beta
\equiv 8\pi P/B^2$, 
which is closely related to the ratio of the sound speed $v_S$ to the
Alfv\'{e}n speed $v_A$; $v_S^2/v_A^2 = \gamma\beta/2$, where $\gamma$ is the
ratio of specific heats. If $\beta$ is either very small or very large, the
acoustic and Alfv\'{e}n frequencies are very different, with consequences 
for choice of simulation
technique.

 The ratios of the mean flow speed $v$ to $v_S$ and
$v_A$, respectively, are denoted by the sonic and Alfv\'{e}n Mach numbers $M$
and $M_A$. When $M\gg 1$, compressibility effects are important and the flow
is pervaded by shocks, requiring an algorithm which treats them accurately. 

The importance of self gravity is measured by the Jeans length $\lambda_J$.
Turbulent pressure effectively increases $\lambda_J$.
This is captured by a heuristic expression for $\lambda_J$ in a
turbulent
medium
\begin{equation}\label{lambdaJ}
\lambda_J\equiv
\left[\frac{\pi v_S^2\left(1+M^2(\lambda_J)\right)}{G\rho}\right]^{1/2},
\end{equation}
where $M^2(\lambda_J)$ is the Mach number of the turbulence at wavelengths less
that $\lambda_J$; turbulence at longer wavelengths does not contribute to
pressure support \cite{CHA1951,BFH1987,PUD1990}.
Equation (\ref{lambdaJ}) reverts to the usual definition
of the Jeans length as $M\rightarrow 0$.
Truelove et al. \cite{TKM1997} have shown that unless the size of the grid, $\Delta
x$, remains less than about $0.25$ of the local $\lambda_J$, the system is
subject to spurious local fragmentation.

\subsection{Microscopic Parameters}

Complete expressions for plasma transport 
coefficients are given by \cite{BRA1965}.
We are primarily concerned here with the magnetic and viscous
diffusivities, which are conveniently given in terms of the
electron
and ion collision times $\tau_e$ and $\tau_i$ 
\begin{equation}\label{tauei}
\tau_e = \frac{2.9\times 10^{-2}}{(\Lambda/10)}\frac{T^{3/2}}{n_eZ}s;\;\;
\tau_i = \frac{1.7}{(\Lambda/10)}\frac{A^{1/2}T^{3/2}}{n_eZ^3}s.
\end{equation}
In eqn. (\ref{tauei}),
$T$ is given in degrees K, $\Lambda$ is the Coulomb logarithm
($\Lambda = 9.4 - 1.15\log{n_e}+3.45\log{T}$ for $T < 5.8\times
10^5K$ and $\Lambda = 15.9 - 1.15\log{n_e} + 2.3\log{T}$ for $T > 5.8\times
10^5K$),
and $A$
and $Z$ are the ionic atomic number and charge, respectively. The electron
density $n_e$ is expressed in cgs units; cm$^{-3}$. 

The
magnetic diffusivity $\lambda_{\Omega}$ introduced in eqn. (\ref{E}) 
can also be expressed in terms of the
electron skin depth $\delta_e\equiv c(m_e/4\pi n_e e^2)^{1/2}$ = $5.4\times
10^{5}n_e^{-1/2}$cm and $\tau_e$ as
\begin{equation}\label{lOhm}
\lambda_{\Omega}=\frac{\delta_e^2}{\tau_e}=9.9\times 10^{12}\left(\Lambda/10
\right)\frac{Z}{T^{3/2}}cm^2s^{-1}.
\end{equation}
This numerical expression allows us to estimate
the Lundquist number $S$
\begin{equation}\label{S}
S\equiv\frac{L_Bv_A}{\lambda_{\Omega}}=0.022\frac{L_BT^{3/2}B}{n_e^{1/2}
(\Lambda/10)},
\end{equation}
where the last expression holds in a hydrogen plasma. Under
astrophysical conditions, $S$ is
enormous: for example, in ionized interstellar gas with $T=10^4$, $n_e =
1$, $B=3\times 10^{-6} G$, and $L_B=1$ pc, $S=10^{17}$. With the exception
of extremely dense, cold environments such as protostellar disks, or
systems with extremely small lengthscales, such as the outer layers of
accreting neutron stars, 
$\lambda_{\Omega}$ is always much less than the numerical diffusivity
arising from discretization.

The viscous diffusivity, or kinematic viscosity $\nu$, is
\begin{equation}\label{nu}
\nu = \frac{kT}{m_i}\tau_i=\frac{1.4\times 10^8}{(\Lambda/10)}\frac{T^{5/2}}{n_eA^{1/2}Z^3}cm^2s^{-1}.
\end{equation}
The Reynolds number $R$ with respect to an organized flow $V$ on a 
lengthscale $L$ is defined as $LV/\nu$, and is almost always extremely large,
confirming our expectation that astrophysical flow can be structured over a wide
range of scales.

According to eqns. (\ref{lOhm}) and (\ref{nu}), the ratio of viscous to
magnetic diffusivity, which is known as the magnetic Prandtl number $P_r$, is
\begin{equation}\label{Pr}
P_r\equiv\frac{\nu}{\lambda_{\Omega}}=\frac{1.4\times 10^{-5}}{(\Lambda/10)^2}\frac{T^4}{n_eA^{1/2}Z^4}.
\end{equation}
Typically, $P_r\gg1$; in the example of fully
ionized interstellar gas introduced above, $P_r=10^{11}$. Therefore,
the kinetic energy spectrum should be truncated on much larger scales than
the magnetic fluctuation spectrum \cite{KUA1992,MAC2002,CLV2002} 
The computations by Maron \& Cowley \cite{MAC2002}
show that the essentially constant shear of the velocity field on
scales near the resistive scale generates tightly folded, or hairpin-like,
magnetic structures. 

If the ion cyclotron frequency $\omega_{ci} \sim 1.8\times 10^4BZ/A$ 
and the ion collision time
$\tau_i$ satisfy the condition $\omega_{ci}\tau_i\gg 1$, 
the viscosity is highly 
anisotropic with respect to the magnetic field. The viscous
force acting on shear flow parallel to the magnetic field is reduced by a
factor of $(\omega_{ci}\tau_i)^2$, while the parallel viscosity remains the
same (see \cite{BRA1965} for a full description). Using
eqn. (\ref{tauei}), we see that $\omega_{ci}\tau_i$ is indeed large; in our numerical example, it is
nearly 10$^5$, implying very strong suppression of viscosity perpendicular
to the magnetic field. Maron \& Cowley \cite{MAC2002} have implemented the full tensor
viscosity in their simulations of turbulent dynamos.

\subsection{Implications for Numerical Techniques\label{subsec:numimplications}}

In order for a numerical method to include diffusive processes, we would ideally require
two things, namely (a) that it can resolve the physically important scales and
(b) that it can handle the disparate time scales.
Clearly, meeting each of these requirements alone is already a non-trivial task.

Resolving the physically interesting scales means separating the intrinsic numerical
diffusion existing in any scheme from its physical counterpart. Simultaneously, it is
desirable to cover scales much larger than the diffusive scales as well, in order
to embed the problem in a realistic environment. Currently, studies are restricted
to one of these regimes, for obvious reasons.
 
There are various ways to parametrize the small-scale physics in the large scale 
approach. Shock capturing is a familiar example (for a detailed discussion see e.g. \cite{LMD1998}).
Many finite-difference codes apply a von-Neumann-Richtmyer artificial viscosity
\cite{NER1950}, spreading out the shock across several support points and not affecting
the solution in smooth regions. Artificial viscosity 
mimicks physical viscosity in that it satisfies the 
Rankine-Hugoniot conditions, but on a much larger scale, which
renders the numerical scheme very robust at low computational cost. It 
operates only where the flow is compressive.

Galsgaard \& Nordlund \cite{GAN1995} implemented an artificial resistivity algorithm 
similar to artifical viscosity in their simulations
of formation and dissipation of current sheets in the solar corona; see also
Caunt \& Korpi \cite{CAK2001}. 

``Current sheet capturing" is intrinsically more complex than shock
capturing.
As we discuss in \S 4.3,
the energy dissipated by a shock is independent of the magnitude of the
diffusivity,
but this does not hold for a current sheet. Moreover, the topological 
evolution of the magnetic field is determined 
primarily by processes in current sheets, and this imposes additional
requirements on how they are treated.
Detailed, small scale
studies could be useful for developing a parameterization of the effects of
diffusion on energy balance and topology. This is reminiscent of the approach
taken in large eddy simulations, which numerically resolve the largest scales
while treating the unresolved scales with a subgrid model
(e.g. \cite{CAN1994,COO1999,ROC2001}).

Godunov-type methods \cite{GOD1959} avoid broadening the shocks over many zones.
They solve the Riemann problem defined by a set 
of advection equations for the physical variables and the physical states on both sides of the 
discontinuity. The 
upstream and downstream states are connected by a sequence of shocks or
rarefaction waves, each of which satisfies the Rankine-Hugoniot conditions. 
As implementation of a full Riemann solver 
would involve determining the full wave structure and all wave
propagation speeds, a variety of approximate Riemann solvers has been
designed to simplify the process by disregarding certain wave types or linearizing
the problem (see \cite{LMD1998} for details). 
Riemann solvers utilize the wave nature of hyperbolic equations, however, the
diffusive terms are parabolic, so that they cannot be included in the 
Riemann solver directly.

An alternative approach has been taken in the so-called BGK-schemes \cite{BGK1954}, using
the collisionless Boltzmann equation as a model of gas dynamics (see \cite{PRX1993}, and
\cite{TAX2000} for MHD).

Needless to say, the $P_r\gg 1$ regime cannot be captured by numerical codes
which solve the ideal fluid equations ($\eta=\nu=0$), since the numerical
diffusivities for magnetic field and velocity are about the same size and occur
at about the same scale, i.e. the grid scale. 
This is an argument for including viscosity and resistivity in the equations, in 
which case the physical diffusivities must be larger than their numerical counterparts,
and their scales must be well separated. 
This is the approach taken by Maron \& Cowley \cite{MAC2002}, who
used a spectral code to study turbulence with widely separated, but resolved,
viscous and resistive scales.

Several efforts have been made to connect the spatial scales by adaptive
mesh refinement techniques \cite{BEO1984,BEC1989}
for large eddy simulations \cite{COO1999}, 
general hyperbolic systems \cite{BER1998},
incompressible non-ideal MHD \cite{FGM1997} 
and for compressible MHD \cite{ZIE2001,BAL2001}.
However, turbulent flows by definition connect scales over the whole
domain, which complicates finding a suitable refinement criterion and may in
the end yield only modest savings of time. 

We have seen in \S\ref{subsubsec:macroparm} that timescales in astrophysical problems can 
differ by several orders of magnitude. Including diffusive processes aggravates the problem,
as their time scales are generally orders of magnitude longer than the dynamical times. Any disparity in timescales leads to a stiff problem.
Explicit methods -- often the first choice because of their low time and memory needs --
advance the solution by a time step $\Delta t$ which is given by the so-called
Courant - Friedrichs - Levy (CFL) condition \cite{CFL1928}
\begin{equation}
  \Delta t \leq \frac{\Delta x}{c}
\end{equation}
with the distance between two support points $\Delta x$ and a characteristic propagation speed $c$.
Disparate characteristic speeds then lead to severe time step restrictions. 
Moreover, the slowly varying components may introduce numerical errors \cite{HUR2001}.

One solution to the problem is to choose a short timestep for the fast
processes, and update the variables controlled by slow processes less
frequently. Mac Low
et al. \cite{MNK1995}
used this so-called subcycling in their 
implementation of ambipolar diffusion in the ZEUS code.
Depending on the problem, the critical processes often can be broken out and treated implicitly, 
while the remaining equations can be treated explicitly \cite{MAS1997}. 
Fully implicit schemes (e.g. \cite{JSE1997}) can be unconditionally stable and independent of the 
time step chosen. Their main limitation lies in the tremendous computational needs in the
multi-dimensional case. 
For a discussion of implicit methods
and for a comparison between explicit and implicit shock capturing methods see 
\cite{HUR2001,YEE1997}. For a discussion of higher order finite difference
schemes and their application to MHD turbulence simulations see \cite{BRA2001}.

As this brief discussion shows, a great variety of techniques can be brought
to bear on MHD turbulence problems. Which technique is best depends very
much on the problem at hand.

\section{Results from Theories and Simulations}

Theory offers an interpretive framework for simulations, while numerical
experiments test turbulence theory, and can inspire new developments.
Here, we discuss amplification of a field by turbulence in the context of dynamo
theory, magnetic fluctuations, the formation of current sheets, and
the dynamical effects of turbulence on large scales. The first three of these
problems are closely related. The fourth is topical in view of recent simulations of star 
forming regions.

\subsection{Turbulent Amplification of a Weak Field\label{subsec:turbamp}}

Amplification of a weak magnetic field by turbulence is one of the main
components of dynamo theory. 
A successful astrophysical dynamo theory must demonstrate that a dominant,
large scale magnetic field can be generated by small scale turbulence,
on a timescale which is virtually independent of the Lundquist
number $S$ in the limit $S\rightarrow\infty$. This is a highly nonlinear
problem, which involves multiple scales. It probably cannot be solved without
numerical simulations, but in view of the extreme parameters involved it is
unlikely to yield to brute force alone.

The growth rate of magnetic energy in a periodic domain is given by eqn.
(\ref{Wdot}). Although the resistive term appears as a sink, some resistivity
is necessary to effect irreversible topological change. Amplification occurs
through work done by the flow. In general, there is also a surface integral
$-c\int\bd\bS\cdot\bE\times\bB$ on the RHS of eqn. (\ref{Wdot}), which 
represents an energy flux through the boundaries.

Equation (\ref{Wdot}) says nothing about the structure of the field, and, in
particular, whether magnetic energy is concentrated at large or small scales.
If we formally set $\lambda_{\Omega}\equiv 0$,
the ratio of fieldstrength to line length is
preserved by incompressible motions. The only way to lengthen fieldlines in a
finite volume is to wind or tangle them. Therefore, we expect that most of
the energy will be in the \textit{rms} field, rather than in the mean field.
For example, if the protogalactic magnetic
field were 10$^{-17}$ G \cite{GFZ2000},
the fieldlines would have been lengthened by more
than a factor of 10$^{11}$ in the course of amplifying the field to 
its present strength. Yet, the large scale and small scale components of the
galactic magnetic field are observed to be roughly equal \cite{ZWH1997}.
It is not enough to argue that magnetic forces will prevent the field
from becoming tangled on small scales. If the lines aren't
somehow lengthened, the
field won't be strengthened.

Of course, we are interested in $\lambda_{\Omega}\rightarrow 0$, not
$\lambda_{\Omega}\equiv 0$. The fieldlines are allowed to break, and the
question is whether they do so in a way which prevents power from piling up
at the resistive scale and peaking instead at the large scale.
Because $S$ is so large, topological constraints are strong, and we
expect resistive boundary layers to form. 

Let's see how theories and simulations of dynamo theory address this problem.
The most influential and detailed dynamo theory is the so-called mean field
theory, which was proposed by Parker \cite{PAR1955}, 
and extended and formalized by Steenbeck, Krause \& R\"adler \cite{SKR1966}. 
Mean field theory is based on the idea that a large
scale magnetic field $\langle\bB\rangle$ can be generated 
from the average inductive electric field $\langle\bE\rangle$ associated with 
small scale, helical velocity and magnetic fluctuations
$\bdelta\bv$, $\bdelta\bB$. The dynamo property of small scale, helical
fluctuations is known as the $\alpha$ effect. There is also diffusive transport 
of $\langle\bB\rangle$, which is known as the $\beta$ effect. Although $\alpha$
and $\beta$ are in general tensors, it suffices for our purposes to write them
as scalars. The induction equation for the mean field is
\begin{equation}\label{meanind}
\frac{\partial\langle\bB\rangle}{\partial t}=-c\bnabla\times\langle\bE\rangle,
\end{equation}
where
\begin{equation}\label{alpha}
\langle c\bE\rangle +\langle\bV\rangle\times\langle\bB\rangle=\lambda_{\Omega}
\bnabla\times\langle\bB\rangle -
\langle\bdelta\bv\times\bdelta\bB\rangle\equiv\alpha
\langle\bB\rangle + \left(\lambda_{\Omega}+\beta\right)
\bnabla\times\langle\bB\rangle + ....
\end{equation}
In eqn. (\ref{alpha}),
$\langle\bV\rangle$ is the large scale velocity field, and the ``..." represent
additional terms involving moments of the turbulent fluctuations and
successively higher derivatives of $\langle\bB\rangle$ (see Moffatt \cite{MOF1978}). Note 
the correspondence between eqns. (\ref{meanind}) and (\ref{alpha}) and eqns. (\ref{faraday}) 
and (\ref{E}). 

If $\alpha$ and $\beta$ are independent of $\langle\bB\rangle$, eqn (\ref{meanind}) is 
linear in  $\langle\bB\rangle$. Physically, this corresponds to neglecting the back 
reaction of the magnetic field on the turbulent velocity, so the linear case is sometimes 
called the kinematic case.

The linear version of eqn. (\ref{meanind}) has 
solutions which vary exponentially in time. 
At large shear and low wavenumber $k$, the growth
rate is approximately the geometric
mean of the shear rate $V^{\prime}$ and the turbulent frequency $k\alpha$.

Standard mean field dynamo theory does not directly address the evolution of
the small scale field, and the role of resistivity, on which the fate of the
small scale field depends, is left somewhat implicit. Usually 
$\lambda_{\Omega}$ is dropped from eqn. (\ref{meanind}), since $\beta$ is
assumed to be much larger. A calculation by Moffatt \cite{MOF1978} of the $\alpha$ effect
due to helical Alfv\'{e}n waves demonstrates that $\alpha\equiv 0$
unless there is a phase difference between $\bdelta\bv$ and $\bdelta\bB$. 
In Moffatt's
model, the phase lag arises from magnetic diffusivity.
At small $\lambda_{\Omega}$, 
$\alpha\propto\lambda_{\Omega}$, implying slow dynamo action at large
$R_m$.

In more general calculations of $\alpha$, the diffusive damping time $(k^2
\lambda_{\Omega})^{-1}$ is replaced by the correlation time $\tau$ of the
turbulence. At large $R_m$, this is much shorter than the resistive time, and
implies large growth rates for the dynamo. But $\beta$ does not distinguish
between converting large scale field to small scale field and actually
destroying field, so the behavior of the 
small scale fields are obscured by this treatment.
Kulsrud \& Anderson \cite{KUA1992} used the tools of mean field theory itself to
follow the growth of the small scale field. They
showed that at large $R_m$, the rms field $\langle B^2\rangle^{1/2}$ grows
much faster than the mean field $\langle\bB\rangle$, and is dominated by small
scale fluctuations with a $k^{3/2}$ power spectrum. 

A completely different perspective on kinematic dynamo theory emerges from
solutions of the full magnetic induction equation for prescribed, chaotic
flow. These studies, which are fully described up to
1995 in the volume by Childress \&
Gilbert \cite{CHG1995}, use a combination of numerical simulations at large $R_m$
and analytical theory to establish connections between the properties of the
flow and the properties of the magnetic field. Although there are known
examples of flows which amplify the magnetic field at a rate independent of
$R_m$, as originally shown by Galloway \& Proctor \cite{GAP1992}, 
the fields produced by these so-called fast
dynamos are highly intermittent in space and
fluctuate in sign, confirming at least qualitatively that small scale fields
are dominant in kinematic dynamos at high $R_m$.

The kinematic theories are modified by dynamical effects.
It was expected on general grounds that the dynamo would saturate as the
magnetic field approached equipartition with the kinetic energy. A
saturation process known as ``Alfv\'{e}nization" was first
described by (Pouquet et al \cite{PFL1976}). Saturation occurs because
 the small scale fluctuations increasingly resemble Alfv\'{e}n waves as the mean
field grows. In 
an ideal medium $\bdelta\bB\rightarrow\pm\bdelta\bv$ in Alfv\'{e}n
units, so their cross product also tends to zero, eliminating the $\alpha$
effect (also shown in the calculation by Moffatt).
Alfv\'{e}nization was first identified
by Pouquet et al. \cite{PFL1976} using a spectral closure scheme, and later derived
using quasilinear theory by Gruzinov \& Diamond \cite{GRD1994}. In the model of
Pouquet et al., kinetic helicity injected at the forcing scale generates
magnetic helicity, which undergoes both a direct cascade, to high $k$, 
where it is resistively dissipated, and an
inverse cascade, to low $k$. Alfv\'{e}nization saturates the dynamo once the
large scale field is roughly in equipartition with the kinetic energy. In the
model of Gruzinov \& Diamond, saturation by Alfv\'{e}nization takes place when the
ratio of kinetic to large scale magnetic energy is $\mathcal{O}(R_m)$,
confirming the dominance of small scale fields at large $R_m$. The difference
between these calculations may lie in their treatments of diffusion. 
Gruzinov \& Diamond explicitly include $\lambda_{\Omega}$, while Pouquet et al.
took all the diffusivities to be eddy diffusivities, which are much
larger than molecular diffusivities.

The role of kinetic helicity injection has been tested in simulations.
Computations by Maron \& Cowley~\cite{MAC2002}, without kinetic helicity injection, show a
peak in spectral power at the resistive scale, as predicted by Kulsrud \&
Anderson~\cite{KUA1992}. Simulations by Maron \& Blackman~\cite{MAB2002} show that a
large scale field is generated when kinetic helicity
is injected, but that the growth rate is proportional to the resistive
time. Nonlinear simulations with helical forcing by Brummell et al. 
\cite{BCT2001}
show that the magnetic lengthscale is proportional to $R_m^{-1/2}$, while
\textit{kinematic} investigations testing the sensitivity of fast dynamo
action to kinetic helicity have not found an inverse cascade ~\cite{HCK1996}.
 
Thus, we see that much hinges on the resistive time, or, more broadly, the
mechanism by which small scale fields are dissipated. Turbulent diffusion of
$\langle\bB\rangle$, parameterized by the $\beta$ effect, merely stands for
spectral transfer out of the large scale. The possibility remains that 
turbulence could mix the field to the small scale, efficiently destroying it.
However, numerical
models by Cattaneo et al. \cite{CHK1996}, and subsequent analytical
work by Kim \cite{K2000} suggest that the mixing is self limiting due to Lorentz
forces. 

The possibility that the small scale fields might escape through
an open boundary rather than being destroyed \textit{in situ}, 
leading to growth of the large scale field on the
convective timescale rather than the resistive timescale, was raised by Parker
\cite{PAR1992}  and Blackman \& Field \cite{BF2000} 
Up to now, neither analytical studies \cite{RK2000} nor
numerical models \cite{BRD2001} have confirmed this idea.

Astrophysical dynamo theory is at a critical juncture. The
fundamental premise of 
mean field dynamo theory - that kinetic helicity injected at small scales
can drive the growth of a magnetic field at large scales - is supported
by a variety of analytical treatments and numerical investigations.
Mean field theory
is a felicitous outcome of turbulence theory in the sense that it can be
applied just by calculating a few parameters (or functions): $\alpha$, $\beta$,
and the large scale shear. However, neither theory nor simulations
have fully come to grips with the large value of $R_m$ and its implications
for the small scale fields. From a numerical point of view, computations
at large $R_m$ which can capture current sheets as well as accommodating
a variety of boundary conditions would seem to be a prerequisite for either
validating or superseding mean field theory.

\subsection{Magnetic Fluctuations}

Steady 
hydrodynamic turbulence, uninfluenced by  boundaries,
 has been characterized in two regimes. 
Subsonic turbulence is incompressible, and follows the Kolmogorov scaling ($E(k)
\propto k^{-5/3}$). Supersonic hydrodynamic turbulence is compressible, and
better described as Burgers turbulence ($E(k)\propto k^{-2}$). (These power
law spectra, and all subsequent ones, describe only the so-called
inertial range, in which there is no driving or microscopic dissipation, only
nonlinear spectral transfer.)

Both these limiting forms of turbulence are spatially intermittent in the sense
that the variances of quantities such as the rate of energy dissipation undergo
substantial fluctuations from point to point. In Burgers turbulence the basic
structures are shock waves, and the large fluctuations and gradients occur
principally in the shock fronts. In Kolmogorov turbulence, the basic units are
eddies, and intermittency is represented by the concentration of vorticity
into small structures. Intermittency does not affect the energy spectrum of 
Burgers turbulence 
(the Fourier spectrum of a shock itself is $k^{-2}$), but is thought to steepen the
Kolmogorov spectrum. This occurs through the local enhancement of the
dissipation rate at small scales \cite{LES1990}. 

Hydrodynamic turbulence in unstratified, nonrotating systems is isotropic. A
uniform component of magnetic field causes turbulence to be anisotropic.

An ideal, adiabatic fluid with a uniform component of magnetic field has three
linear modes: the shear Alfv\'{e}n mode, and the fast and slow magnetosonic modes.
The velocity field $\bdelta\bv$ of the Alfv\'{e}n mode, is solenoidal, and this
mode is driven purely by magnetic tension. The dispersion relation is
$\omega = k_{\parallel}v_A\equiv\omega_A$,
where $k_{\parallel}$ is the component of $k$ parallel to the large scale
magnetic field. The fast and slow magnetosonic modes are compressive, and
driven by a combination of gas pressure, magnetic pressure, and magnetic
tension.

Since the shear Alfv\'{e}n mode is noncompressive, it is reasonable to ask whether
a strong ($\beta < 1$) magnetic field permits a regime of pure shear, 
supersonic turbulence. The
answer is no. 

First, suppose $\bk_{\perp}\equiv 0$. Then,
the nonlinear magnetic pressure gradient associated with the
fluctuating transverse magnetic field drives a compressive parallel flow,
causing an Alfv\'{e}n wavetrain to steepen nonlinearly into a train of weak shocks
\cite{COK1974}, much as a sound wave steepens. The only exception is 
the infinitely long, circularly polarized wave, which is an exact
solution of the ideal MHD equations. The outcome of the evolution of
an ensemble of parallel propagating Alfv\'{e}n waves is a series
of shocks, similar to Burgers turbulence, with $k^{-2}$ kinetic and magnetic
energy spectra \cite{GAO1996}.
The steepening time depends on amplitude as $(\delta v/v_A
)^{-2}$, while the steepening time of an acoustic wave is proportional to
$(\delta v/v_S)^{-1}$. In this sense, shear waves survive longer in a low
$\beta$ plasma than compressive waves of the same amplitude, by a factor of
order $\beta^{-1/2}(v_A/\delta v)$, but the outcome in either case is a
compressive flow dominated by shocks, and the difference in timescales is only
significant for small wave amplitude.

Although Burgers turbulence was originally based on 1D flow, 3D simulations
of supersonic MHD turbulence show similar rapid evolution to a shock
dominated flow. The distribution of shock strengths in 
simulations of driven and decaying
turbulence, with and without magnetic fields, has been studied by Smith et al. 
\cite{SMZ2000,SMH2000}.

The steepening of Alfv\'{e}n waves, which can be viewed as a cascade in 
$k_{\parallel}$, is suppressed if $\beta\ge 1$. Under the assumption of 
incompressibility,  and in contrast to the
highly compressible low $\beta$ case, interactions occur only between
oppositely directed wave packets. In particular there is no self 
interaction of the kind which leads to steepening. 

The incompressible
shear Alfv\'{e}n wave cascade is highly 
anisotropic, with power transferred much more rapidly in $k_{\perp}$ than in
$k_{\parallel}$. In physical space, the correlation length transverse to the
mean field is much shorter than the correlation length along it. These
anisotropic states can be described by so-called reduced MHD  \cite{S1976},
which is a quasi-2D approximation 
describing elongated magnetic structures. It is often applied to
(and, indeed, was derived for), low $\beta$ plasmas, but it neglects 
the parallel steepening which leads to shocks. This, as we
said above, is a good approximation
for long parallel Alfv\'{e}n transit times and small turbulent amplitudes. 

There is not yet complete consensus on the cascade itself. Weak
MHD turbulence, in which an individual Alfv\'{e}n wave packet 
survives for many periods, can be viewed as  weakly perturbative resonant
interactions between multiple waves. Sridhar \& Goldreich \cite{SRG1994} argued that the
dominant interaction is a four wave interaction, which results in a spectrum
$E(k)\propto k^{-7/3}$. Others \cite{NGB1996,BHN2001,GNN2002}
claim that the dominant interactions are three wave 
interactions, and predict $E(k)\propto k^{-2}$. 

In strong MHD turbulence, wave packets survive for only about one wave period.
Goldreich \& Sridhar \cite{GOS1995} developed the concept of a critically balanced cascade
in which the wave frequency $\omega_A$ is the same as the nonlinear frequency
$k_{\perp}v_{\perp}$. This, together with the
requirement of constant spectral
energy flux argument $k_{\perp}v_{\perp}^3$, leads to a
Kolmogorov spectrum $E(k)\propto k^{-5/3}$ and wavenumber anisotropy
$k_{\parallel}/k_{\perp}\propto k_{\perp}^{-1/3}$. Other theories of strong
MHD turbulence predict $E(k)\propto k^{-3/2}$, including the original isotropic 
Iroshnikov-Kraichnan theory \cite{NAK2001}.

Numerical simulations, if free of confounding computational effects, could
play a role in resolving these disagreements. Biskamp \& M\"uller~\cite{BIM2000}
and Cho \& Vishniac \cite{CHV2000} find $E(k)\propto k^{-5/3}$ over about 
one decade in $k$ space. Maron \& Goldreich \cite{MAG2001} slightly extended the inertial
range and found $E(k)\propto k^{-3/2}$. These authors argue that the
spectrum is flattened because of intermittency. The crucial difference
between the HD and MHD cases is that energy cascades in MHD only when
oppositely directed wave packets collide; the small filling factor
associated with intermittency reduces the collision rate, more than offsetting
the enhanced dissipation associated with small structures. 

Larger computations, with an extended inertial range, should shortly become
available. The differences in the spectra obtained with different codes
and under different forms of driving (Biskamp \& M\"uller studied decaying or
isotropically forced turbulence, Cho \& Vishniac studied isotropically forced
turbulence, and Maron \& Goldreich studied anisotropically forced turbulence)
are at this point comparable to the theoretical disagreements. A joint
exercise in which different groups 
simulated identical problems and implemented
identical diagnostics might be quite enlightening.

A variety of processes can terminate the cascade at short wavelengths (see
\cite{LIG2001} for a recent discussion). These include fluid effects,
such as ion-neutral friction, and kinetic effects such as ion gyroresonance
absorption and electron Landau damping \cite{QG1999}. The extension
of the magnetic spectrum to scales below the viscous cutoff in a high Prandtl
number plasma is discussed by \cite{MAC2002}
and \cite{CLV2002}.

\subsection{Intermittency: Current Sheets and Flux Tubes}

Simulations of MHD turbulence show strong intermittency in the distribution
of magnetic field gradients, or current. In the weak field, high Prandtl
number simulations of \cite{MAC2002}, the fluctuating field is tightly
folded. These results suggest that we look at current sheets.

Current sheets and filaments 
are sites of intense local heating, and
possibly particle acceleration, which makes them interesting
in their own right. It is quite likely that in most astrophysical systems,
significant departures from
flux freezing occur only in these regions, so their existence is important in
the evolution of magnetic field topology, and for dynamo action, 
(see \S\ref{subsec:turbamp}).
Reconnection at magnetic X-points drives strong, small scale jets. 
Simulations by Galsgaard \& Nordlund~\cite{GAN1996} show highly time variable 
dissipation rates associated with the formation and disruption of large scale current 
sheets.

Current sheets are a prerequisite of, and accompany, magnetic reconnection.
Reconnection is a resistive process which occurs on a 
timescale intermediate
between the Ohmic time and the Alfv\'{e}n time. The reconnection time
$\tau_{rec}$ is
often parameterized by the Lundquist
number $S$; $\tau_{rec}$ scales as $S^p$ with $0 \le p < 1$. If $p
= 0$, the reconnection is said to be fast (in correspondence with the definition of a fast
dynamo). In Sweet-Parker reconnection, $p=1/2$
\cite{P1979}, and resistive tearing
modes in slabs and cylinders have $p=3/5$, and $p=1/3$, respectively
\cite{B1978}.
The formation and behavior of current sheets has been studied for many years.
Most of this work has focussed on plasmas in or near magnetostatic
equilibrium, which can be expected if $\beta\ll 1$ and the plasma is either
allowed to relax or is driven at frequencies far below its characteristic
Alfv\'{e}n frequency. Such conditions hold, for example, in stellar coronae, in
which the magnetic fieldlines are tied to the underlying photosphere, and
evolve quasistatically in response to slow photospheric motion.

Current sheets form as a plasma seeks equilibrium while obeying the topological
constraints imposed by flux freezing \cite{M1985}.
Topological constraints can
arise from features such as null points \cite{S1971}
or separatrices \cite{LOW1988,ZP1990,LC1996}. Sweet ~\cite{S1958} and 
Parker ~\cite{P1972} suggested that 
shearing and winding the footpoints of an initially uniform field will create
current sheets.
Even a simple sinusoidal deformation of the boundaries of a sheared
magnetic slab can induce the formation of a current sheet \cite{HK1985}.

As a simple example of how current sheets can form, consider the
magnetic field of a pair of neighboring, uniformly magnetized superconducting
spheres, with antiparallel magnetic dipole moments, surrounded by a perfectly 
conducting, zero pressure plasma. Some fieldlines have both
endpoints on the same sphere, and others have one endpoint on each sphere.

Now suppose that one sphere rotates on its magnetic axis by an angle $\phi$.
The fieldlines which are not connected to the other sphere also rotate by
$\phi$, but the fieldlines which are connected to the other sphere cannot
rotate uniformly because one endpoint remains fixed. The 
separatrix surface which divides 
the two domains of magnetic connectivity becomes a current sheet:
the field outside it is sheared while the field inside it is not.

There is still no comprehensive theory for exactly when current sheets form, and
how their properties reflect the underlying magnetic topology. Nevertheless,
analytical arguments \cite{V1985,LOS1994} and numerical 
experiments \cite{MSV1989,LOS1994,GAN1995}
show that shear, or equivalently current density, grows exponentially
rapidly in line tied magnetic fields with randomly moving footpoints. This has
some of the same consequences as an exact singularity, which cannot form in
any case in even a slightly resistive plasma.

These results are not fully applicable to MHD turbulence, in which the
magnetic field is not in equilibrium. Nevertheless, current
sheets form in flows with magnetic null points \cite{P1987}.
The underlying mechanism for the growth of the current is
the random stretching of neighboring fieldlines, which leads to
large cross-field shear.

What are the consequences of current sheet formation for dissipation of
magnetic energy? That is, if 
energy per unit volume $\dot e$ is being added to a
system by a driving process, under what circumstances can this energy
can be dissipated in current sheets?

Consider a current sheet of area $L^2$
and width $\delta$. Assume the current sheet structure is given by
the Sweet-Parker theory, so that
$\delta=LS^{-1/2}=(L\lambda_{\Omega}/v_A)^{1/2}$. If $N(L)dL$ is the number
density of current sheets with $L$ between $L$ and $L+dL$, then the rate at
which energy is dissipated by current sheets in this size range is
\begin{equation}\label{dotdw}
d\dot w=\frac{B^2}{4\pi}\frac{v_A}{L}N(L)L^2\delta dL = \frac{B^2}{4\pi}N(L)
\left(L^3\lambda_{\Omega}v_A\right)^{1/2}dL.
\end{equation}
The rate of energy dissipation by all current sheets is obtained by integrating
eqn. (\ref{dotdw}) over $L$:
\begin{equation}
\dot w =\frac{B^2}{4\pi}
\left(\lambda_{\Omega}v_A\right)^{1/2}\int dLN(L)L^{3/2}.
\end{equation}
If $\dot e$ is independent of $\lambda_{\Omega}$ as $\lambda_{\Omega}
\rightarrow 0$, then $\int dLN(L)L^{3/2}$ must scale as $\lambda_{\Omega}^{-1/2}$ in this limit. This would imply an unbounded number of current sheets 
as $\lambda_{\Omega}\rightarrow 0$, which is virtually equivalent to a pileup
of magnetic energy in fluctuations at the resistive scale.
\footnote{If we make a similar argument for the efficiency of energy dissipation in
shocks, the result is independent of viscosity, because the shock thickness
is directly proportional to viscosity. Therefore, the inviscid limit does
not require an infinite number of shocks to dissipate the energy input by
driving.}

There are alternative possibilities. One is that the current sheets are not
structured according to the Sweet-Parker picture. If $\delta$ scaled as
$\lambda_{\Omega}$, as it does (up to a logarithmic factor) in Petschek's
model of fast reconnection
(reviewed in \cite{P1979}), then $\int dLN(L)L^{3/2}$ could be independent of
$\lambda_{\Omega}$. Note, however, that while in the Sweet-Parker theory,
magnetic energy is converted to kinetic energy and to heat at about the same
rate, in Petschek's theory magnetic energy is converted predominantly to
kinetic energy, and that the realizability of Petschek's model has
recently been
questioned \cite{B1993,UZK2000}. 

The second possibility is that the magnetic and
velocity fields adjust themselves such that the driving just balances the
damping, as they do in mechanical systems such as linear
harmonic oscillators. Thus, 
the energy input rate $\dot e$ would depend on $\lambda_{\Omega}$, while the
applied force itself, presumably would not. 

Dedicated simulations of current sheets and reconnection regions
themselves allow us to study the small scale aspects of the problem. 
It is equally important to
understand how current sheets are embedded in the overall flow. This has already been done to some 
extent for laboratory experiments \cite{UZK2000}.
Ultimately, it may be possible to parameterize magnetic reconnection in
simulations on larger scales.

Finally, we mention an even more extreme form of intermittency: the segregation
of the magnetic field into thin tubes. This is observed in the
solar photosphere, where magnetic flux tubes, or sheets, just a few hundred
km thick are seen at the
borders of convective cells. Flux tubes can form only in high $\beta$ plasma,
so that gas pressure can balance magnetic pressure at the tube walls.

Early attempts to explain photospheric flux tubes relied on a two stage
process, proposed by Parker, for concentrating a diffuse field into organized
structures. According to this scenario, thermal convection would sweep the
field to the borders of convective cells, concentrating the field up to
equipartition with the flow. Then reduced convective heat transport would
cool the gas, ``collapsing" the tube and bringing it to equipartition with
the thermal plasma. 

An alternative possibility is that intermittent fields
are generated \textit{in situ}.
It has been shown that  small scale turbulent
thermal convection operates as a dynamo, and generates a magnetic field with
a broad tail of high energy features \cite{CAT1999}.

It is unclear whether a dynamo operating in a high $\beta$ plasma should
always produce flux tubes. In the solar case, the turbulence is organized by
stratification and the thermal gradient, and is subsonic. If the turbulence
were supersonic and the field were amplified to equipartition, then 
flux tubes could not persist.

\subsection{Dynamical Effects of Turbulence}

The stresses associated with turbulence can exert dynamical forces. In MHD
turbulence, these forces include both magnetic stresses and the Reynolds
stress. Observations suggest that these stresses are important in the
interstellar medium, and especially in molecular clouds.

It was realized early that 
supersonic motions are nearly ubiquitous in
molecular gas, and pose a problem. If the motions
are associated with gravitational collapse, the implied star formation
rate would be very high \cite{ZUE1974}, 
but if the motions are turbulent, they should be
dissipated very quickly \cite{GOK1974}. This conundrum, together with
the expectation that reasonably strong magnetic fields should be present,
prompted analytical studies of Alfv\'{e}nic turbulence under molecular cloud
conditions \cite{ARM1975,ZWJ1983,E1985}.
This work demonstrated that even when $M_A\ll M$ and the waves are purely transverse,
ion-neutral friction and nonlinear steepening are strong damping
mechanisms, which limit the lifetimes of Alfv\'{e}n waves to a few 10$^6$ yr or
less under molecular cloud conditions.

Even so, under the assumption that the waves might be replenished by some
energy source, various aspects of the theory of self gravitating clouds 
supported by Alfv\'{e}n wave stresses were developed and applied to model clouds 
\cite{PUD1990,FAA1993,MCZ1995,MHP1997,MCH1999}.
According to this theory, which is based on the the weak turbulence
approach pioneered by Dewar \cite{DEW1970}, Alfv\'{e}n waves have an isotropic stress tensor
$\bP=P_w\bI$. The waves
pressure - density relation is $P_w\propto\rho^{1/2}$ in a
stratified, static medium, while $P_w\propto\rho^{3/2}$ under slow,
spatially uniform changes in density with time. The negative polytropic
index ($n=-1/2$) of the $P - \rho$ relation leads to a large center to
surface pressure contrast, in accord with observations (see \cite{MCH1999} for
a good discussion).

The theory of wave supported clouds is most relevant to objects which are
slightly magnetically supercritical, that is, the ratio of mass to 
magnetic flux is slightly too large for magnetostatic support. If the clouds
were subcritical they would be supported by the DC magnetic field. If
they were highly supercritical, the amplitude of the turbulence required to
sustain them would be so large that the weak turbulence theory would
probably not apply, although they could still be turbulently supported.
>From an observational viewpoint, the most likely candidates for wave
support are dense cores with sonic or mildly supersonic linewidths 
\cite{MF1992}.

The picture of strongly supersonic turbulence which emerges from numerical
simulations of molecular clouds 
is quite different from this analytical picture, primarily
because, as we discussed in \S 4.2,
 generation of compressive disturbances is entirely unavoidable.
These compressive disturbances, or shocks, sweep up most of the mass into
thin layers in which the local Jeans length is small. These layers, and the
even denser structures formed where layers intersect, collapse if
the domain is
magnetically supercritical, whether the
turbulence is freely decaying or maintained in a steady state by driving
\cite{HMK2001,OSG2001}. The driven models are globally
stable in the sense that $\lambda_J$ computed using the total turbulent
energy and mean density is larger than the length of the domain. The small,
high density structures collapse because the turbulent power at wavelengths
less than the local Jeans length is relatively small (see
eqn. (\ref{lambdaJ})), and because the
intensity of turbulence inside the cores is comparable with the density
outside. Nevertheless, the bound structures are small, and the role of
numerical diffusion within them, which damps the turbulence and removes
the support by the DC magnetic field, has not been quantified. Therefore,
while the formation of turbulently supported clumps has not been found in
the numerical models, their existence cannot be ruled out based on the present
models. 

\section{Ambipolar Drift}

As we have discussed in the preceding sections, both theoretical and
computational aspects of
astrophysical magnetohydrodynamics
are dogged persistently by the enormous value of $R_m$. 
There is one situation,
however, in which the magnetic diffusivity is large whether or not it is 
enhanced by turbulence: 
in weakly ionized gas, the 
magnetic field and plasma
drift with respect to the neutrals. This so-called ambipolar drift is not
entirely
equivalent to resistive diffusion, because it preserves magnetic topology (the
field is frozen to the plasma), but it does change the ratio of magnetic flux
to mass, as originally pointed out in \cite{MS1956},
and dissipates fluctuations on small scales. 

Full treatment of ambipolar drift requires solving the equations of MHD for
the plasma and neutrals treated as distinct species coupled by collisions,
ionization, and recombination \cite{DRA1986}. At the low ionization fractions expected in
molecular clouds, the ion Alfv\'{e}n speed $v_{Ai}\equiv
B/\sqrt{4\pi\rho_i}$ can be several
orders of magnitude larger than the other characteristic speeds in the problem,
reducing the maximum possible timestep by a similar factor. This and other
numerical issues  related to implementation of ambipolar drift
are discussed by \cite{MNK1995,TOT1995,MAS1997,STO1997}. 

On timescales longer than the ion - neutral collision time
$\tau_{in}$, and at low
ionized mass fractions, 
the ion-neutral drift $v_D$ 
is determined by 
balancing the Lorentz force on the ions against the frictional drag 
by neutrals
\begin{equation}\label{vD}
\bv_D=\frac{\left(\bnabla\times\bB\right)\times\bB}{4\pi\rho_i\nu_{in}}\equiv\frac{v_{Ai}^2
\tau_{in}}{L_B}\hat f,
\end{equation}
where $\hat f$ is a unit vector in the direction of the Lorentz force.
The second equality in eqn. (\ref{vD})
can be taken as a definition of the magnetic lengthscale $L_B$. The
quantity $v_{Ai}^2\tau_{in}$ 
has units of diffusivity, and from now on we
refer to it as $\lambda_{AD}$. With this definition, $v_D=\lambda_{AD}/L_B$.
Note that $\rho_i/\tau_{in}=\rho_n/\tau_{ni}$ and $\rho_n\approx\rho$, so
$v_{Ai}^2\tau_{in}=v_{A}^2\tau_{ni}$, and $\lambda_{AD}$ is often expressed in
the latter terms.

The ambipolar Reynolds number $R_{AD}$ is defined as the ratio of the bulk
velocity $v$ to the drift velocity $v_D$
\begin{equation}\label{RAD}
R_{AD}\equiv\frac{L_Bv}{\lambda_{AD}},
\end{equation}
and measures how well the magnetic field is frozen to the bulk fluid on
scale $L_B$ \cite{ZWB1997}. Setting $R_{AD}=1$ correctly predicts the
thickness of shocks in which the main dissipation mechanism is ion-neutral
friction \cite{DRM1993}, the wavelength at which Alfv\'{e}n waves are
critically damped \cite{KUP1969}, and the minimum size of an
eddy which can wind up a magnetic field \cite{ZWB1997}.

Simulations of supersonic MHD turbulence \cite{PZN2000} with 
varying strengths of ambipolar drift shows that, on average, the drift
smooths out the current, and reduces the rms Lorentz force.

We expect (with one caveat discussed at the end of this section) that the
magnetic field should show very little structure below the scale $L_{min}
\equiv \lambda_{AD}/v$ at which
$R_{AD} = 1$. Consider a volume $L^3$ of 
turbulent gas with mean density $\rho$, turbulent velocity
$v$, and mean magnetic field $B$. If we introduce the crossing time $\tau_d
\equiv L/v$ and the Alfv\'{e}n Mach number $M_A\equiv v/v_A$, then we can write
%
\begin{equation}\label{lmin}
\frac{L_{min}}{L}=\frac{1}{M_A^2}\frac{\tau_{ni}}{\tau_d}.
\end{equation}
%
%

Expressing $v$ in units of km s$^{-1}$ as $v_5$, and $L$ in pc as $L_{pc}$,
$\tau_d\sim 3\times 10^{13}L_{pc}/v_5$ s.
For the particle species expected
in molecular clouds, $\tau_{ni}\sim 5
\times 10^8 n_i^{-1}$s.
With the ionization equilibrium relation $n_i\sim 10^{-5}n_n^{1/2}$ (cgs;
\cite{MZG1993}), eqn. (\ref{lmin}) becomes
\begin{equation}\label{lminnum}
\frac{L_{min}}{L}=1.7\frac{v_5}{n^{1/2}L_{pc}}\frac{1}{M_A^2}.
\end{equation}
Comparing eqn. (\ref{lminnum}) with the grid spacing $\Delta x/L=1/N$, we
see that a moderately large numerical
simulation (say 256$^3$), which is not highly super-Alfv\'{e}nic or extremely
dense,
should resolve almost all the magnetic structure associated with turbulence.

Equation (\ref{lmin}) can also be written in terms of the
critical magnetic field $B_c\equiv 2\pi G^{1/2}\rho L$ and the
free fall time $\tau_{ff}\equiv (4\pi G\rho)^{-1/2}\sim 5.5\times 10^{14}
n_n^{-1/2}$s as
\begin{equation}\label{lminalt}
\frac{L_{min}}{L}=\left(\frac{B}{B_c}\right)^2\frac{\tau_{ni}\tau_{d}}{4\tau_{ff}^2}\sim\left(\frac{B}{B_c}\right)^2\frac{\tau_{ni}}{4\tau_{ff}},
\end{equation}
where the last step holds for virial equilibrium; $\tau_{ff}\sim\tau_d$.
Of course, eqn. (\ref{lmin}) is entirely independent of $G$. 

The modest value of $R_{AD}$ in molecular clouds has been commented upon
and exploited by \cite{KHM2000,OSM2002}. 

Ambipolar drift is essentially diffusive when the magnetic field has a well
ordered, well combed structure. Because the ambipolar diffusivity is
proportional to $B^2$, the diffusion is nonlinear. Sharp fronts can be 
generated in the vicinity of magnetic nulls, and minima are steepened
\cite{BRZ1994,MNK1995,ZWB1997,MAS1997}.

\section{Summary and Future Agenda}

The theme of this paper is how numerical simulations can best cope with the
enormous magnetic Reynolds numbers encountered in most astrophysical problems.
We argued that, because $R_m$ (and the Lundquist number $S$) are so large,
resistive effects occur primarily in thin sheets or filaments, which must be
treated accurately in order to follow the topological evolution of the
magnetic field.

In \S 2, we reviewed the magnetic induction equation, giving the solution in
the ideal limit, and derived equations for magnetic helicity and magnetic 
energy. We tested helicity conservation in the ZEUS code by simulating the
evolution of the ideal kink mode, and studied fluctuations of helicity in a
simulation of a turbulent cloud. The first of these problems is especially
suitable for benchmarking numerical codes. That is, it yields a 
quantitative estimate
of the numerical resistivity in the code, from which one can compute the
resistive timescale (at least in smooth regions, on large scales) and assess
whether it is much longer than other timescales of interest. One can also
compare the performance of different codes.

In \S 3, we discussed some important dimensionless parameters in astrophysical
MHD problems. We verified the large size of the Lundquist number $S$ and
magnetic Reynolds number $R_m$ under most conditions, and showed that the
magnetic Prandtl number, the ratio of viscous to magnetic diffusivity, is
also usually large. This implies that magnetic structure will generally extend
to smaller scales than velocity structure. We briefly summarized a variety of
numerical techniques from the perspective of the large $S$ limit. 

In \S 4, we discussed theoretical and numerical results on several basic
problems. In \S 4.1 we discussed the amplification of magnetic fields by
turbulence, and argued that the successful operation of a large scale dynamo
depends on fast diffusion of the magnetic field. In \S 4.2 we discussed MHD
turbulence itself. In \S 4.3 we discussed current sheets, which may provide
the fast diffusion necessary for a dynamo, and may be produced in turbulence
as a form of intermittency. In \S 4.4, we discussed the dynamical effects of
turbulence, and its importance in molecular clouds.

In \S 5, we discussed ambipolar drift as a mechanism for increasing the
magnetic diffusivity in low density, weakly ionized gas. It is now possible
to simulate turbulent molecular clouds at a 
resolution which accounts for all magnetic structure down to the ambipolar
diffusion scale. Under certain conditions, however, ambipolar drift can
mediate the formation of current sheets, possibly leading to rapid magnetic
reconnection.

It is fortunate that a
number of different groups are simulating MHD turbulence
under astrophysical conditions, using a variety of codes and techniques. There
is substantial agreement on a number of issues, such as the short decay time
of supersonic turbulence, but disagreement on others, such as the spectrum of
strong MHD turbulence. A dedicated effort at benchmarking, in which different
groups simulated identical problems and implemented identical diagnostics,
would provide some perspective. We suggested in \S 4.2 that such an exercise
be carried out for strong MHD turbulence.

It is unlikely that it will ever be possible to adequately resolve current
sheets and global dynamics in a single simulation. Local studies of current
sheets and magnetic reconnection are essential in capturing the physics of
these layers. It is equally important to understand how they are
embedded in the overall
flow, as this constrains their properties, sets the boundary conditions for
reconnection, and is necessary for understanding how to parameterize the
effects of current sheets in global simulations. The inability to deal 
adequately
with this small scale structure is the greatest present challenge to
understanding magnetic field evolution in turbulent flows.

\vspace{0.2cm}

\emph{Acknowledgements:} We are grateful to E. Falgarone and T. Passot for
organizing the conference, and for their hospitality. Our discussions with
A. Burkert, M.-M. Mac Low, and J. Maron were especially useful, as were 
L. Mestel's comments on the manuscript.
FH gratefully acknowledges support by the Feodor-Lynen
program of the Alexander von Humboldt Foundation. The (U.S.) National Science
Foundation provided partial support through grants AST-9800616 and AST-0098701
to the University of Colorado, and through support to the National Center for
Atmospheric Research.


\end{document}